\documentstyle[12pt,epsf,epsfig]{article}

\newlength{\dinwidth}
\newlength{\dinmargin}
\def\lapproxeq{\lower .7ex\hbox{$\;\stackrel{\textstyle
<}{\sim}\;$}}
\def\gapproxeq{\lower .7ex\hbox{$\;\stackrel{\textstyle
>}{\sim}\;$}}
\def\be{\begin{equation}}
\def\ee{\end{equation}}
\def\bea{\begin{eqnarray}}
\def\eea{\end{eqnarray}}

\makeatletter
\def\fmslash{\@ifnextchar[{\fmsl@sh}{\fmsl@sh[0mu]}}
\def\fmsl@sh[#1]#2{%
\mathchoice
{\@fmsl@sh\displaystyle{#1}{#2}}%
{\@fmsl@sh\textstyle{#1}{#2}}%
{\@fmsl@sh\scriptstyle{#1}{#2}}%
{\@fmsl@sh\scriptscriptstyle{#1}{#2}}}
\def\@fmsl@sh#1#2#3{\m@th\ooalign{$\hfil#1\mkern#2/\hfil$\crcr$#1
#3$}}
\makeatother

\begin{document}
\titlepage
\begin{flushright}
DTP/97/100 \\
November 1997 \\
\vspace*{1in}
\end{flushright}
\renewcommand{\thefootnote}{\fnsymbol{footnote}}
\begin{center}
{\large \bf The effect of off-diagonal parton distributions in  \\
diffractive vector meson electroproduction} \\
\vspace*{0.5in}
A.D.\ Martin and M.G.\ Ryskin\footnote[1]{Petersburg Nuclear
Physics Institute, SU-188 350 Gatchina, Russia.} \\
Department of Physics, University of Durham, DH1 3LE, UK
\end{center}
\renewcommand{\thefootnote}{\arabic{footnote}}

\vspace*{2cm}

\begin{abstract}
We present a simple physical description of the off-diagonal gluon
distribution $g (x, x^\prime, Q^2)$ and splitting function 
$P_{gg} (x, x^\prime)$ which enter the amplitude
for diffractive vector meson production.  We study the off-diagonal effects
both on the evolution in $Q^2$ and on the input distribution.  We predict
the ratio $R$ of the off-diagonal to the diagonal gluon density, 
$x^\prime g (x, x^\prime, 
Q^2) / x g (x, Q^2)$, as a function of the kinematic variables.
\end{abstract}
\newpage

\noindent{\large \bf 1.~Introduction} \\

It is well known that the cross section of hard scattering processes 
(such as deep inelastic scattering, the production of large $p_T$ jets, etc.) 
can be written as the sum of parton distributions multiplied by the cross 
sections of hard subprocesses calculated at the parton level using perturbative 
QCD.  That is we can factor off the long distance (non-perturbative) effects 
into universal, process independent, parton distributions specific to the 
incoming hadron.

Strictly speaking a parton distribution is the diagonal element of a density 
matrix, or the diagonal element of an operator $\hat{O}$ in the Wilson 
operator product expansion (OPE).  To be more specific let us start with the 
Fock decomposition of the wave function of an incoming hadron, say a proton
\be
\label{eq:a1}
| p \rangle = \sum_k \; \psi_k (\mbox{\boldmath$a$}_1, \ldots 
\mbox{\boldmath$a$}_n)
\ee
where $k = qqq, qqqg, \ldots$ and $n$ is the number of partons in the Fock 
state $\psi_k$.  Each parton is described by a vector $\mbox{\boldmath$a$}_i$ 
with coordinates which specify its momentum, colour, spin, flavour etc.  To 
calculate an inclusive cross section we must sum over the coordinates of all 
the partons except for the one which participates in the hard subprocess.  
Thus, for example, the gluon distribution is the diagonal element
\be
\label{eq:a2}
g (x) \; = \; \langle p | G G | p \rangle \; = \;  \sum_k \int | \psi_k 
(\mbox{\boldmath$a$}_1, \ldots \mbox{\boldmath$a$}_n) |^2 \;  
\prod_{i \neq g}^n \; d \mbox{\boldmath$a$}_i.
\ee
where as usual the operator product $GG$ denotes the product of an 
annihilation operator $G$ and a creation operator $G^{\dagger}$.  The 
product \lq \lq counts" the number of gluons in the Fock states.  Contrary 
to the conventional notation the operator $G$ corresponds to a gluon  with 
fixed momentum fraction $x$, not to a point $\mbox{\boldmath$r$}$ in 
coordinate space.

On the other hand there are a set of reactions which are described by 
the off-diagonal elements of the density matrix, say $\langle N^{\prime} | 
\hat{O} | N \rangle$ where the momentum, helicity and/or charge of the outgoing nucleon $N^{\prime}$ are not the same as those of the ingoing nucleon $N$.  Examples are :
\begin{itemize}
\item[(a)] virtual photon Compton scattering $(\gamma^* p \rightarrow \gamma p)$ 
or vector particle electroproduction $(\gamma^* p \rightarrow Vp)$ --- examples 
of the latter processes include $\gamma^* p \rightarrow \rho p$, $\gamma^* p 
\rightarrow Z p$, $\gamma^* p \rightarrow W^+ n$ and $\gamma^* p \rightarrow 
(\mu^+ \mu^-) p$;
\item[(b)] processes which involve the BFKL amplitude at large 
$|t| \:  = \: | p - p^{\prime} |^2 \neq 0$;
\item[(c)] the polarized structure function $g_2$, or alternatively
the function $g_{\bot} = g_1 + g_2$  
for transversely polarised nucleons.  
\end{itemize}
In each case we have to study the evolution of the matrix element of the two 
gluon operator $GG$ (or the two quark operator $q \overline{q}$) evaluated 
between {\it non-diagonal} states, $\mbox{\boldmath$a$}_i \neq 
\mbox{\boldmath$a$}_i^{\prime}$.  In case (a), $\gamma^* p \rightarrow V p$, the 
important difference in $\langle p^{\prime} | G G | p \rangle$ is between the 
longitudinal components of the incoming and outgoing proton momentum, where
\be
\label{eq:a3}
t_{{\rm min}} \; = \; (p - p^{\prime})_{\parallel}^2 \; = \; -  \; \left(
\frac{M_V^2 + Q^2}{W^2} \right)^2 \; m_p^2.
\ee
$Q^2$ is the virtuality of the photon, $W$ is the $\gamma^* p$ centre of mass 
energy and $M_V$ is the mass of the produced vector meson.  On the other hand 
in case (b) the large difference is due to the transverse components
\be
\label{eq:a4} 
|t| \approx (p - p^{\prime})_T^2,
\ee
whereas in case (c) the difference is due to the nucleon helicities, 
$\langle \lambda^{\prime} | \overline{q} q | \lambda  \rangle$ with 
$\lambda \neq \lambda^{\prime}$.

To describe these processes we need to consider off-diagonal elements of the 
density matrix
\be
\label{eq:a5}
\langle N^{\prime} | \hat{O} | N \rangle \; = \; \sum_k \int \; \psi_k ^* 
(\mbox{\boldmath$a$}_1^{\prime}, \ldots , \mbox{\boldmath$a$}_n^{\prime} ) \; 
\hat{O} \psi_k
(\mbox{\boldmath$a$}_1, \ldots, \mbox{\boldmath$a$}_n) \; \prod _{i \neq g}^n 
\; \delta (\mbox{\boldmath$a$}_i - \mbox{\boldmath$a$}_i^{\prime}) \; 
d \mbox{\boldmath$a$}_i \: d \mbox{\boldmath$a$}_i^{\prime}
\ee
where the coordinates of the Fock state $\psi$ and $\psi^*$ are all equal 
$(\mbox{\boldmath$a$}_i = \mbox{\boldmath$a$}_i^{\prime})$ except for the 
parton, say $i  = g$, which participates in the hard scattering.  A topical 
example is the off-diagonal gluon distribution $g(x, x^{\prime}) = \langle
p^{\prime} | GG | p \rangle$ for which 
the longitudinal momentum fraction carried by the gluon $(i = g)$ differs on 
the left-$(\psi^*)$ and right-$(\psi)$ hand sides of the amplitude, $x_g \neq 
x_g^{\prime}$. 

In this paper we concentrate on just such off-diagonal parton effects which
occur in the description of diffractive vector meson production, since these
processes appear to be the most accessible experimentally.  We introduce the 
appropriate variables in Section~2.  Off-diagonal effects can change both the
evolution and the input distributions.  We study off-diagonal evolution in 
Section~3.  Although the off-diagonal splitting functions are already known,
this section contains a particularly simple and physical derivation of 
$P_{gg} (x, x^\prime)$.  The off-diagonal effects on the starting distributions
(which are used for the evolution) are discussed in Section~4.  In Section~5
we present numerical estimates of the size of these off-diagonal effects as
functions of $x, \; x^\prime$ and $Q^2$.  Finally in Section~6
we present our main conclusions. \\

\noindent{\large \bf 2.~Diffractive electroproduction of vector mesons} \\

Perhaps the most relevant reactions, at present, are the quasielastic 
diffractive electroproduction of vector mesons, $\gamma^* p \rightarrow V p$.  
These processes are now being studied with increasing precision at HERA.  It 
has been argued \cite{GLUON,RRML} that the amplitudes for these reactions are, to a 
good approximation, proportional to the gluon distribution, 
$g (x, \overline{Q}^2)$, that is the conventional (diagonal) parton 
distribution measured 
in the global analyses of data for deep inelastic and related hard scattering 
processes.  To lowest order the $\gamma^* p  \rightarrow V p$ amplitude can be 
factored into the  product of the $\gamma^* \rightarrow q \overline{q}$ 
transition, the diffractive scattering of the $q \overline{q}$ system on the 
proton via two-gluon exchange, and finally the formation of the vector meson 
$V$ from the outgoing  $q \overline{q}$ pair.  The crucial observation is that, 
at high c.m. energy $W$, the scattering on the proton occurs over a much 
shorter timescale 
than the $\gamma \rightarrow q \overline{q}$ fluctuation or $V$ meson formation
times.  Observation of $\gamma^* p \rightarrow V p$ at c.m. energy $W$ therefore 
probes the gluon distribution $g (x, \overline{Q}^2)$ with
\be
\label{eq:a6}
x \; = \; \frac{Q^2 + M_V^2}{W^2}, \quad \quad \overline{Q}^2 = z (1 - z) 
\: Q^2 + k_T^2 + m_q^2
\ee
where $z$, $\mbox{\boldmath$k$}_T$ and $1 - z$, $- \mbox{\boldmath$k$}_T$ are 
the $q, \overline{q}$ longitudinal momenta fractions and transverse momenta 
with respect to the incoming photon, and $m_q$ is the mass of the quark. Indeed
if this simple model can be theoretically substantiated, the dependence of the 
cross section on the {\it square} of the gluon density means that data for
$\gamma^* p \rightarrow Vp$ (and for the photoproduction of heavy vector mesons
e.g. $\gamma p \rightarrow J/\psi p)$ offer an especially sensitive probe of
the gluon.

However we have noted that $\gamma^* p \rightarrow V p$ is actually described 
by a non-diagonal distribution $g (x, x^{\prime})$ and so it is important to 
quantify\footnote{Recent work on this subject can also be found in ref.\ 
\cite{FFGS}.} the difference between this improved description and the 
approximation of using the conventional, diagonal, gluon distribution.  To be 
specific let us study $\gamma^* p \rightarrow \rho p$ at high $Q^2$.  Due to 
the proton form factor, the transverse momentum of the $\rho$ meson is strongly 
limited $(p_T^{\prime 2} \ll Q^2)$ and the difference between the amplitude 
and the one with diagonal gluons arises predominantly from the transfer of the 
longitudinal momentum which is needed to put  the $\rho$ meson on-mass-shell.  
The partonic decomposition of the process is sketched in Fig.~1.  Also shown 
are the relevant kinematic variables.

Consider the imaginary part of the amplitude, which is the dominant 
component.  Then the intermediate particles are on-mass-shell as indicated by 
the crosses on the lines in Fig.~1.  In order to put the quarks $(k$ 
and $\hat{k})$ on-shell we require the fraction of the proton's 
longitudinal momentum that is carried by the gluon to be
\be
\label{eq:a7}
x \; = \; \frac{Q^2 + M_{q \overline{q}}^2}{W^2}.
\ee
$M_{q \overline{q}}$, the mass of the $q \overline{q}$ pair, is given by 
\be
\label{eq:a8}
M_{q \overline{q}}^2 \; = \; (k + \hat{k})^2 \; \simeq \;
\frac{k_T^2 + m_q^2}{z} \; + \; \frac{\hat{k}_T^2 + m_q^2}{1 - z} \; - \;
l_T^2
\ee
where recall that $(z, \mbox{\boldmath$k$}_T)$ and $(1 - z,  
\mbox{\boldmath$\hat{k}$}_T \: = \: \mbox{\boldmath$l$}_T
- \mbox{\boldmath$k$}_T)$ specify the $q$, $\overline{q}$ kinematics with 
respect to the photon.

On the other hand for the $q \overline{q} \rightarrow \rho$ transition the 
difference between the mass of the vector meson $M_V$ and the $q \overline{q}$
mass $M_{q \overline{q}}$ 
is very small.  Now in the leading log approximation (LLA) the transverse 
momenta are strongly ordered, and in particular $l_T^2 \ll Q^2$.  Therefore
\be
\label{eq:a9} 
x^{\prime} \; = \; \frac{M_{q \overline{q}}^2 - M_V^2}{W^2} \: \ll \: x.
\ee
Thus we have a longitudinal momentum difference 
\be
\label{eq:a10}
\delta x \; = \; x_i - x_i^{\prime} \; \simeq \; \frac{Q^2 + M_V^2}{W^2},
\ee
which persists along the whole ladder, see Fig.~1. 

The variable $x^\prime$ is not directly measurable.  Since the mass of the $q
\overline{q}$ system, $M_{q \overline{q}}$, is not fixed, the value of
$x^\prime$ is smeared out by the integration over the quark momenta.  It 
means that observation of diffractive vector meson production, 
$\gamma^* p \rightarrow V p$, measures the off-diagonal
gluon distribution over some interval of $x^\prime$.  However this is not such
a disadvantage as it seems.  To leading log accuracy the range of $x^\prime$ is
much less than $x$ and we may safely put $x^\prime = 0$ (see \cite{RRML}).

We wish to study the $\ln Q^2$ evolution of the off-diagonal parton distributions 
indicated by the ladder structure in Fig.~1. First we note that this is part of 
a more general evolution.  In fact for the electroproduction of a massive vector
boson, say $\gamma^* p \rightarrow Zp$, the off-diagonal evolution equation
\cite{DMRGH,BB,R,JI,R2} describes not only the deep inelastic evolution for spacelike
$Q^2$, which is our concern here, but also the Brodsky-Lepage \cite{BL} 
leading $\ln M_V^2$ evolution of the vector boson wave function which occurs in
the timelike $Q^2$ region.\footnote{This is analogous to the Gribov-Lipatov
\cite{GL} relation which connects the splitting functions for $ep \rightarrow eX$ deep
inelastic evolution in the spacelike $Q^2$ region with the splitting functions
which occur in $e^+ e^-$ annihilation (or, to be more precise, in $e^+ e^-
\rightarrow q X$) at moderate $x$ in the timelike $Q^2$ region.} For instance
for heavy vector boson production $\gamma^* p \rightarrow V p$ we have evolution
arising from the sum of $(\alpha_S \ln M_V^2)^n$ terms in the region 
of $x^{\prime} < 0$.  Since now both the 
$x$ and $x^{\prime}$ gluons go in the same direction, we are in the
timelike region.  However,
we lose at least one $\ln Q^2$ at the cell where $x^{\prime}$ changes sign. 
When $x^{\prime}$ goes negative the parton reverses direction.  Therefore
there are no collinear logs at this point and, moreover, $x^{\prime} < 0$ 
kinematics is inconsistent with causality in the LLA.  Indeed the $x^{\prime} 
< 0$ domain corresponds to the decay of the heavy vector boson into the 
virtual photon and two gluons, for example $Z \rightarrow \gamma^* gg$.  In the LLA it is 
described by the point-like production of two massless (on-shell) gluons.  
However these two gluons cannot interact through $s$ channel gluon exchange 
(as shown in Fig.~1): the $x^{\prime} < 0$ gluon has insufficient time to 
overtake the gluon $x$.  A more formal and detailed discussion is given by 
Radyushkin \cite{R}.  So at leading order we must terminate the evolution at 
$x^{\prime} \simeq 0$ and the amplitude of the process sketched in Fig.~1 is 
proportional to the off-diagonal gluon distribution\footnote{We use a factor
$x^{\prime}$, rather than $x$, to avoid the $1/x^{\prime}$ singularity for 
$x^{\prime} \ll x$. It is informative to amplify this comment. Recall that for
the diagonal gluon density we use $x g (x)$ to avoid the Bremsstrahlung 
singularity, where $g (x) dx$ describes the probability to find a gluon in an
interval $dx$ of momentum space.  When we express observables in terms of the
gluon distribution we find that the relevant region of the $dx$ integration
is of the order of $x$.  It would therefore have been more natural to introduce
parton densities $x g (x)$ in the place of $g (x)$. In the off-diagonal case, the
region of the $dx$ integration (which corresponds to $d \mbox{\boldmath$a$}_g$ in
(\ref{eq:a2})) is limited by minimum $\{ x, x^{\prime} \} = x^{\prime}$.  So the
appropriate density is $x^{\prime} g (x, x^{\prime})$.} $x^{\prime} g
(x, x^{\prime}) $ with
\be
\label{eq:a11}
x \; \simeq \; \frac{Q^2 + M_V^2}{W^2}, \quad x^{\prime} \simeq 0
\ee

In the $Q^2$ evolution we keep the full $x$ dependence. For the process under 
consideration, $\gamma^* p \rightarrow \rho p$ at 
HERA energies, the values of $x$ are very small, $x \lapproxeq 10^{-3}$.  
In this small $x$ region the leading contribution has a double logarithmic
form in which at each iteration we have not only $\ln Q^2$ but also a
$\ln 1/x$ arising from the integration over longitudinal momenta.  This 
double leading logarithmic part of the amplitude,\footnote{Recall that the double logs come
from the integration over the rapidity of the $s$ channel gluon (and from the
integral over the strongly ordered transverse momenta).  In the notation of
ref.\ \cite{RRML} the upper limit in the $d x_s / x_s$ integration is given by 
the largest of the $t$ channel momentum fractions in the $i^{{\rm th}}$ cell, 
that is by $x_i$ rather than by $x_i^{\prime}$.  Hence we find $\ln 1/x$ and
not $\ln 1/x^{\prime}$.}
\be
\label{eq:a12}
x^{\prime} g (x, x^{\prime}) \; \simeq \; \sum_n \;  C_n (\alpha_S \ln 
\frac{1}{x} \; \ln Q^2)^n,
\ee
corresponds to the domain in which the longitudinal, as well as transverse, 
components are strongly ordered, $x_i \ll x_{i - 1}$.  As in this domain
$\hat{x}_i \simeq x_i$, there is no $x^{\prime}$ dependence and the off-diagonal
distribution becomes the diagonal one.\footnote{This was the conclusion of the
first paper \cite{BARTELS} on off-diagonal distributions.}

Therefore we can consider the use of the diagonal parton density as a zero 
approximation and study the difference between the diagonal 
$(\delta x \rightarrow 0)$ and off-diagonal amplitudes.  That is we will 
calculate the ratio
\be
\label{eq:a13}
R \; = \; \frac{x^{\prime} g (x, x^{\prime})}{x g (x)} \quad {\rm where} \quad
x^{\prime} \; = \; x - \delta x.
\ee
There are two possible sources of difference between the diagonal and off-diagonal
densities which causes the ratio $R$ to differ from unity.
First the input distributions at the starting scale $Q_0^2$ of the evolution 
may differ and, second, there is a difference in the evolution itself
on account of the difference
in the splitting functions.  It is convenient to discuss the effect on the 
evolution first. \\

\noindent{\large \bf 3.~Off-diagonal evolution equation and splitting function} \\

The off-diagonal splitting functions which describe the log $Q^2$ evolution of 
partons have been widely discussed for example \cite{GRBH,JI,R,R2,FFGS}. They have
been calculated by the QCD-string operator approach \cite{BB,R} or by using the
conventional LLA in which the partons are put on-shell, that is by the QCD 
extension of the Weizs\"{a}cker-Williams technique\footnote{The splitting functions
were originally discussed in \cite{GLR}, but unfortunately there were mistakes
in the formulae.}. So the off-diagonal splitting functions $P (z, z^{\prime})$
are known.

Since the gluons play a dominant role in the small $x$ HERA domain it
should be a reasonable simplification to consider pure gluon evolution.  
The evolution equation for the 
off-diagonal gluon density is then
\be
\label{eq:a21}
\frac{d g (x, x^{\prime})}{d \ln Q^2} \; = \; \frac{\alpha_S (Q^2)}{2 \pi} \;
\int_0^1 \: \frac{dz}{z} \; P_{gg} (z, z^{\prime}) \: g 
(x/z, x^{\prime}/z^{\prime})
\ee
where in the notation of Fig.~1
\be
\label{eq:a21a}
z^{\prime} \; = \; \frac{x_n^{\prime}}{x_{n - 1}^{\prime}} \; = \; 
\frac{x - \delta x}{x/z - \delta x}.
\ee
Note that in (\ref{eq:a21}) there is no integration over $z^{\prime}$.  To see
why this is so recall that the $d z^{\prime} / z^{\prime}$ integration is equivalent
to the $d x^{\prime} / x^{\prime}$ integration and that $ x^{\prime} \; = \;
x - \delta x$.  Since $\delta x$ is fixed by (\ref{eq:a10}) then $z^{\prime}$ is
fixed also. The splitting kernel is given by \cite{GRBH,JI}
\bea
\label{eq:a22}
P_{gg} (z, z^{\prime}) & = & 2N_C \: z (1 - z^{\prime}) \; 
\left\{ 1 + \frac{1}
{z z^{\prime}} + \frac{1}{(1 - z)(1 - z^{\prime})} + \frac{z - z^{\prime}}
{2 z^{\prime} (1 - z)} + \frac{z^{\prime} - z}
{2 z (1 - z^{\prime})} \right\} \nonumber \\ 
\nonumber \\
& - & (V + V^{\prime}) \: \delta(1 - z).
\eea
The integrals $V$ and $V^{\prime}$ in the virtual term correspond to the left-
and right-hand sides of the amplitude.  We have
\be
\label{eq:a23}
V \; = \; N_C \int_0^1 \: z d z \left[ z (1 - z) + \frac{1}{z} + \frac{1}
{1 - z} \right]
\ee 
In an attempt to give physical insight of the structure of $P_{gg}$ we will, 
below, present a particularly simple derivation of (\ref{eq:a22}).  

First we notice that
if $z = z^\prime$ then $P_{gg} (z, z^\prime)$ reduces to the conventional 
splitting function $P_{gg} (z)$.
Recall that in the conventional (diagonal) case, the $1/(1 - z)$ singularity 
as $z \rightarrow 1$ in the real
part of the kernel (which corresponds to the emission of a soft gluon with 
momentum fraction $y_i = x_i (1 - z))$ is cancelled by the singularity in the 
virtual contribution which is needed to restore the normalization of the parton
wave function.  To be explicit, the real-virtual cancellation is 
\be
\label{eq:a18}
\frac{2 C_2}{1 - z} \; - \; \delta (1 - z) \: \int^1 \: \frac{d \overline{z}}
{1 - \overline{z}} \; 2 C_2,
\ee
which gives a combination that is regular as $z \rightarrow 1$. For the gluon
$C_2 (G) = N_C = 3$. In accordance
with the conventional Bloch-Nordsieck procedure we must cut off the contribution
due to soft gluons with small energies, say with $E \leq E_0$, using the 
{\it same} cut-off $E_0$ in the real and virtual parts of the splitting 
functions.

At first sight it appears from (\ref{eq:a22}) that for off-diagonal evolution
only the pole $1 / (1 - z)$ occurs, but not a $1 / (1 - z^\prime)$ singularity.
On the other hand we see that the normalisation is $2N_C$, which reflects the fact that for
fixed $\delta x = x_i - x_i^\prime$ and $z \rightarrow 1$ (that is $x_i 
\rightarrow x_{i - 1}$) the value of $z^\prime = x_i^\prime / x_{i - 1}^\prime$
also goes to 1. That is the $1 / (1 - z)$ pole represents both the $z \rightarrow
1$ and $z^\prime \rightarrow 1$ singularities.  So the total soft gluon singularity
in the real part with a factor $2N_C$ is 
balanced by the sum of the $V$ and $V^\prime$ virtual terms.  However care is
needed. We note that in the off-diagonal case we have different momentum fractions 
in the \lq \lq left" and \lq \lq right" virtual terms, that is $x \neq 
x^{\prime}$.  Thus we must use slightly different cuts for
\be
\label{eq:a19}
z_i \; = \; x_i / x_{i - 1} \quad {\rm and} \quad z_i^{\prime} \; = \; 
x_i^{\prime} / x_{i - 1}^{\prime}.
\ee
Suppose that we use $1 - z_i > \Delta$ in the virtual term $V$ in the left amplitude,
where $\Delta$ corresponds to the same bound $E > E_0$ that is imposed on the 
real soft gluon emissions, then we must use a cut
\be
\label{eq:a20}
1 - z_i^{\prime} \; > \; \Delta^{\prime} \; = \; \frac{\Delta}
{1 - \delta x/ x_{i - 1}}
\ee
in the virtual term $V^\prime$ in the amplitude on the right. 

It is instructive to sketch the derivation of the off-diagonal splitting
function $P_{gg} (z, z^{\prime})$ of (\ref{eq:a22}).  The relevant triple
gluon vertex is shown in Fig.~2.  To leading order accuracy we can put the gluons
on-mass-shell and then the vertex function
\be
\label{eq:a24}
\Gamma_{\mu \nu \rho} \; = \; \delta_{\mu \nu} (p + k)_\rho \: + \: 
\delta_{\nu \rho}(p - 2k)_\mu \: + \: \delta_{\rho \mu} (k- 2p)_\nu
\ee
reduces to simple forms depending on the directions of the polarisation vectors.
For each gluon we take one of the two independent polarisation vectors to be
$\mbox{\boldmath$\epsilon$}^{(n)}$ in the direction of $\mbox{\boldmath$p$} \times
\mbox{\boldmath$k$}$.  Then the other polarisation vector is taken, for the $s$
channel gluon to be $\mbox{\boldmath$\epsilon$}_\rho^{(t)}$ in the direction 
$\mbox{\boldmath$\epsilon$}^{(n)} \times \mbox{\boldmath$q$}$, and for the $t$
channel gluons to be $\mbox{\boldmath$\epsilon$}_\mu^{(t)}$ and 
$\mbox{\boldmath$\epsilon$}_\nu^{(t)}$ in the directions
$\mbox{\boldmath$\epsilon$}^{(n)} \times \mbox{\boldmath$p$}$ and 
$\mbox{\boldmath$\epsilon$}^{(n)} \times \mbox{\boldmath$k$}$ respectively. It
is convenient to introduce
\be
\label{eq:a25}
\Gamma_{abc} \; = \; \Gamma_{\mu \nu \rho} \; \epsilon_\mu^{(a)} \; \epsilon_\nu^{(b)}
\; \epsilon_\rho^{(c)}
\ee
where $a, \: b, \: c$ must be either $n$ or $t$ to indicate the polarisations
of the three gluons.  Then the only non-zero elements of $\Gamma_{abc}$ are
\bea
\label{eq:a26}
\Gamma_{tnn} & = & - 2k_T, \quad \Gamma_{ntn} \; = \; 2k_T/z, \nonumber \\
\Gamma_{nnt} & = & \frac{2k_T}{1 - z}, \quad \Gamma_{ttt} \; = \; 2k_T
\left(\frac{1}{1 - z} + \frac{1}{z} - 1 \right) 
\eea
where $z = k_{\|}/p_{\|} \: = \: x_i / x_{i - 1}$.  We may make two 
observations.  First we note that (\ref{eq:a26}) satisfies the leading
logarithm approximation requirement that the helicity factor $| \Gamma |^2$ be 
proportional to $k_T^2$.  Second, it is straightforward to see from (\ref{eq:a24}) that
a non-zero $\Gamma_{abc}$ must  have an even number of $n$ subscripts.  We may
use (\ref{eq:a26}) to
obtain the real gluon emission contribution to $P_{gg}$. We sum over the
final, and average over the initial, polarisations, and find
\be
\label{eq:a27}
{\textstyle \frac{1}{2}} \: \sum \Gamma^* (z^\prime) \Gamma (z) \; = \; 4k_T^2
\left\{ 1 \: + \: \frac{1}{z z^\prime} \: + \: \frac{1}{(1 - z) (1 - z^\prime)} 
\: + \: \frac{z - z^\prime}{2z^\prime (1 - z)} \: + \: \frac{z^\prime - z}
{2z (1 - z^\prime)}
\right\}.
\ee
Finally to obtain the contribution to $P_{gg}$ given in (\ref{eq:a22}) we must include the kinematical
factor, $\frac{1}{2} z(1 - z^\prime)$, which is appropriate when writing the 
evolution equation (\ref{eq:a21}) as an integral over $dz/z$.  In the limit
$z^\prime \rightarrow z$ we see that $P_{gg} (z, z^\prime)$ reduces to the 
familiar (diagonal) splitting function $P_{gg} (z)$, and that the last two 
terms in (\ref{eq:a27}) vanish.  These terms come from the cross products in
$\Gamma_{ttt}^* (z^\prime) \; \Gamma_{ttt} (z)$. \\

\newpage

\noindent{\large \bf 4.~Off-diagonal effects on the input distributions} \\
 
Suppose that in the diagonal approximation we provide the input at some (low) 
scale $Q_0^2$.  Now the off-diagonal matrix element can be expressed in terms 
of the Mandelstam variables $(s, \: u, \: t)$ and the masses or virtualities 
of the incoming and outgoing particles.  It corresponds to the elastic parton-nucleon
amplitude, say $A (s, \: t, \: Q_0^2, \: Q_0^{\prime 2})$, which describes the
subprocess shown at the bottom of the ladder diagram of Fig.~1.  We have to 
study two effects.  One due to $t \neq 0$  
and the second due to the difference in the virtualities of 
the two outgoing gluons
\be
\label{eq:a14}
\delta Q^2 \; = \; Q_0^{\prime 2} - Q_0^2 \; \simeq \; Q_0^2 \: \left(\frac{x_0 - 
\delta x}{x_0} \right) - Q_0^2 \; = \; - Q_0^2 \; \frac{\delta x}
{x_0},
\ee
in the notation of Fig.~1. We consider $x \ll 1$.  Of course there are
important off-diagonal effects at large $t$ which originate first from the
form factor of the target proton and second from the BFKL $\ln (1/x)$ evolution
at non-zero $p_T^\prime$.

Here we are not concerned with large $p_T^\prime$ vector meson production.  However
even for zero-angle scattering we still have
\be
\label{eq:a15}
| t | \; = \; | t_{{\rm min}} | \; = \; m_p^2 (\delta x)^2 \; \neq \; 0,
\ee
so we have to consider a potential $t \neq 0$ off-diagonal effect.
Now the scale which controls the $t$ behaviour of the amplitude is given by
the nearest singularity $(t \; = \; 4 m_{\pi}^2)$ or by the slope $B$ in $t$ of 
elastic parton-proton differential cross section $d \sigma / d t \propto
{\rm exp} (B t)$, where the value of $B$ is less than the slope of the elastic
$pp$ cross section  
$B < B_{pp} \simeq 10 {\rm GeV}^{-2}$.  Thus for HERA energies, where $\delta x
\lapproxeq 10^{-2} - 10^{-3}$, the effect coming from $| t_{{\rm min}} | \lapproxeq
10^{-4} {\rm GeV}^2$ is negligible.

On the other hand there may be some suppression of the input distribution due
to $\delta Q^2 \neq 0$.  So far the current \lq \lq off-diagonal" models of the 
nucleon would imply a weak dependence on $\delta x \equiv x - x^\prime$, for $\delta x
\ll 1$, at the initial scale, that is before the evolution is taken into account.
Thus the calculations have taken the input ratio
\be
\label{eq:b1}
R_0 \; = \; \frac{x^\prime g (x, x^\prime, Q_0^2)}{xg (x, Q_0^2)} \; \simeq \; 1
\ee
at $t = t_{{\rm min}}$.  However it is not clear at which value of $Q_0^2$ we
should take $R_0 = 1$.  The evolution causes the ratio $R$ to increase, 
particularly for $x \simeq \delta x$ (that is $x^\prime \simeq 0$). Thus, 
strictly speaking, after we have chosen the starting scale $Q_0^2$, we should
determine $R_0 (x, \delta x)$ by fitting to the data. In other words off-diagonal
evolution requires a two-variable input function $x\prime g (x, x^\prime, Q_0^2)$
at an initial scale $Q_0^2$, which originates from the non-perturbative domain.
Clearly $R_0 (x, \delta x)$ depends on our choice of $Q_0^2$.

Lacking the relevant data to determine $R_0 (x, \delta x)$, the best that we can
do at present is to start the evolution from rather small $Q_0^2$ where
nevertheless we may use perturbative QCD evolution.  At these values of $Q^2$ we
may expect $R_0 \simeq 1$.  The reason is as follows.  On the one hand it appears
unreasonable to have $R < 1$ at small $Q_0^2$, as in conventional phenomenology
the elastic parton (gluon) - proton amplitude $A (s, \: t, \:Q_0^2, \: 
Q_0^{\prime 2})$
generally increases as the virtuality $Q_0^{\prime 2}$ decreases.  On the other
hand we may use the Schwarz inequality
\be
\label{eq:b2}
\int \; \psi^* (x^\prime, \mbox{\boldmath$a$}_i) \: \psi (x, \mbox{\boldmath$a$}_i) \:
d \mbox{\boldmath$\overline{a}$}_g \; \prod_{i \neq g} d \mbox{\boldmath$a$}_i \;
\leq \; \frac{1}{2} \: \int \left(| \psi (x^\prime, \mbox{\boldmath$a$}_i) |^2 \; + \;
| \psi (x, \mbox{\boldmath$a$}_i) |^2 \right) \: d \mbox{\boldmath$\overline{a}$}_g
\: \prod_{i \neq g} \: d \mbox{\boldmath$a$}_i
\ee
to put an upper limit $R_0$(max) on $R_0$.  We are using the notation of 
(\ref{eq:a5}), but for simplicity we do not show the sum $k$ over the Fock states.
By $d \mbox{\boldmath$\overline{a}$}_g$ we mean integration over all the coordinates
of the gluon except for $dx dx^\prime$.  The Schwarz inequality
(\ref{eq:b2}) gives the upper bound
\be
\label{eq:b3}
R_0 \; \leq \; R_0 {\rm(max)} \; = \; \frac{x^\prime g (x^\prime, Q_0^2) \: +
\: x g (x, Q_0^2)}{2 x g (x, Q_0^2)}.
\ee
We notice that in (\ref{eq:b2}), written in terms of Fock wave functions, the
$Q^2$ dependence is not explicit.  The conventional gluon distribution 
$g (x, Q^2)$ represents the number of gluons with transverse momentum satisfying
$k_T^2 \leq Q^2$.  Thus the argument of $Q^2$ plays the role of upper limit of
the $\Pi \: dk_{Ti}^2$ integration in the subspace of the integration over
$\Pi \: d \mbox{\boldmath$a$}_i$.  That is the evolution scale is controlled
by the largest transverse momentum of the partons and not by the virtuality
$Q^{\prime 2}$.  To very good accuracy we find $R_0 {\rm (max)} = 1$, throughout
our $x$ interval of interest, using GRV partons \cite{GRV} at $Q_0^2 = 0.4
{\rm GeV}^2$, whereas for MRS(R2) partons \cite{MRS} we find $R_0 {\rm (max)}
\simeq 1$ in the range $Q_0^2 = 1.3-1.4{\rm GeV}^2$ as $\delta x$ varies from
$10^{-4}$ to $10^{-2}$.

Finally, for completeness, we note that the input for the off-diagonal 
valence quark distribution has recently been discussed in ref. \cite{JMS}.\\

\noindent{\large \bf 5.~Off-diagonal effects from evolution} \\

To indicate the effect of using off-diagonal distributions we calculate
the ratio
\be
\label{eq:b4}
R \; = \; \frac{x^\prime g (x, x^\prime, Q^2)}{x g (x, Q^2)}
\ee
for fixed values of $\delta x = x - x^\prime$, using both the leading order
off-diagonal and the conventional (diagonal) evolution equations for the gluon.
We use the gluon of GRV \cite{GRV} as input at $Q_0^2 = 1.5
{\rm GeV}^2$.  The results for $R (x, \delta x, Q^2)$ are shown in Fig.~3 for four
values of $\delta x$, in each case showing $R$ as a function of $x$ for $Q^2 =
4, \: 20$ and $100 {\rm GeV}^2$.

The main features of the results shown in Fig.~3 are as follows.  First, the
ratio $R$ only noticeably differs from 1 at small $x$, close to $\delta x$.
For $x \gapproxeq 10 \: \delta x$  we see that $R$ is always less than 1.05.
Second, as may be expected, $R$ increases as we evolve up in $Q^2$.  To see the 
reason for the increase we must consider the real and virtual contributions to 
$P_{gg}$.  The diagonal and off-diagonal splitting functions have the same 
$1/z$ behaviour and thus lead to the same double log behaviour,
$\alpha_S \ln Q^2 \: \ln 1/x$, of the real gluon emission contribution.  On the
other hand, if $xg \sim x^{-\lambda} \: (Q^2)^\gamma$, then the anomalous
dimension
\be
\label{eq:b5}
\gamma (\lambda) \; = \; \frac{\alpha_S}{2 \pi} \; \int_0^1 \: P_{gg} (z) \: 
z^\lambda \: dz
\ee
due to the \lq diagonal' $P_{gg}$ is less than that due to the \lq off-diagonal'
$P_{gg}$, mainly due to the larger cut-off $\Delta^\prime$ in the virtual
$V^\prime$ term as compared to $\Delta$, see (\ref{eq:a20}).  We therefore have 
a smaller remainder in the negative $V^\prime$ contribution, after the cancellation
of the singularities as in (\ref{eq:a18}).  This in turn leads to a positive
difference in $\gamma (\lambda)$ and to a faster off-diagonal $Q^2$ evolution.

The third effect apparent in Fig.~3 is the increase of $R$ with $\delta x$. Why
is this?  The difference between the splitting functions $\Delta P \equiv
P_{gg} (z, z^\prime) - P_{gg} (z)$ decreases with decreasing
$z - z^\prime \approx \delta x/x$, see (\ref{eq:a21a}) and (\ref{eq:a22}). 
Thus most of the difference comes from the region of 
evolution where $\delta x \sim x$.  If we evolve from a value of $x = x_0$ which
is much larger than $\delta x$, then it is only the last few evolution iterations
which have $\delta x \sim x$ and so contribute to $R \neq 1$.  From this point of
view we expect that $R$ depends on $\delta x/x$ and not on the absolute value of
$\delta x$.  This is true for very small $\delta x$, and hints of this behaviour
can be seen by comparing the third and fourth plots of Fig.~3, that is those
for $\delta x = 10^{-4}$ and $10^{-5}$
respectively.  However for larger values of $\delta x$, say $\delta x \sim
10^{-2}$ the evolution starts from $x_0 \; (x_0 \sim 0.1)$ which is already
sufficiently close to $\delta x$ so that all intervals of the $Q^2$ evolution
contribute to $R \neq 1$, giving a larger value of $R$.

Fig.~4 shows the effect of taking the input $R_0 = 1$ at a considerably higher
starting scale, namely $Q_0^2 = 4 {\rm GeV}^2$.  The lower two curves in each
plot compare the values of $R(x, \delta x, Q^2)$ obtained using this $Q_0^2$
(dot-dashed curves) with our preferred predictions based on 
$Q_0^2 = 1.5 {\rm GeV}^2$ input (continuous curves).  From the
previous discussion we expect the $Q_0^2 = 1.5 {\rm GeV}^2$ input to yield a 
larger value of $R$ simply because of the longer interval of $Q^2$ evolution.
However we see that the difference disappears during evolution mainly due to the
fact that for very large $Q^2$ the essential contribution to the gluon density
comes from the region of initial $x_0 \gg \delta x$ where the off-diagonal effects
in the initial conditions are negligible.  For interest, we also started the 
evolution from the lowest possible value of $Q_0^2$, namely $Q_0^2 = 0.4
{\rm GeV}^2$, again using the GRV gluon.  The predictions are shown by the 
dashed curves, which by chance coincide, for an appreciable range of $x$,
with the upper limits which we discuss below.

The remaining curve in Fig.~4 is obtained from the upper limit (\ref{eq:b3})
given by the Schwarz inequality.  Of course we may use the inequality at any
$Q^2$ in the perturbative QCD region.  Clearly calculating the bound at the 
lowest reasonable $Q^2$ will give the tightest constraint on $R$.  For this
reason we use GRV gluons to calculate $R$(max) of (\ref{eq:b3}) at
$Q^2 = 1.5 {\rm GeV}^2$ and then evolve both $g (x, x^\prime)$ and $g (x)$
to give the \lq \lq evolved" upper limit shown at the higher $Q^2$ values in
Fig.~4.  This procedure is valid because the evolution equations are linear in
$g$.  For small $Q^2$ the difference between $R$ obtained by evolving from 
the $R_0 = 1$ at $Q_0^2 = 1.5 {\rm GeV}^2$ (and also $Q_0^2 = 4 {\rm GeV}^2)$ and
the upper limit is appreciable.  However as we evolve to large $Q^2$ the system
depends less and less on the initial conditions and the predictions approach
each other.

In Fig.~5 we explore the sensitivity of the predictions for $R$ to the choice
of the input gluon distribution.  We compare the predictions based on using
the gluon from the MRS(R2) set of partons \cite{MRS} with those previously 
obtained using the GRV gluon \cite{GRV}.  In each case we start the evolution
from $Q_0^2 = 1.5 {\rm GeV}^2$.  The predictions are shown by continuous curves,
together with the upper limits obtained by evolving from the Schwarz inequality
evaluated at $Q_0^2 = 1.5 {\rm GeV}^2$. 

We may make two observations concerning Fig.~5. First, we see that the MRS and GRV predictions
for the {\it ratio} essentially coincide at $Q^2 = 4 {\rm GeV}^2$.  However the
individual off-diagonal and diagonal gluons are larger and steeper for GRV than
MRS.  Since $x g \sim x^{- \lambda}$ has a large effective $\lambda$ for GRV
than MRS it gives, on average, a larger $\delta x / x$ and hence a more rapid
off-diagonal evolution.  Thus the GRV prediction for $R$ lies above that of
MRS at $Q^2 = 100 {\rm GeV}^2$.  Second, we see that the MRS predictions lie just
below the upper limit, whereas the upper limit for GRV is considerably higher.
Again this is directly attributable to the steepness in $x$ of the GRV gluons. In
fact the GRV gluons \cite{GRV} are \lq \lq steeper" than that required to fit the new 
high precision measurements of $F_2$ at small $x$.  Therefore the upper limit
based on the GRV gluon should be regarded as too high.  

Qualitatively our results for $R$ have similar features to those given in
\cite{FFGS}.  However our predictions are appreciably smaller.  For example,
at $x = 1.1 \times 10^{-4}$, $x^\prime = 10^{-5}$ and $Q^2 \simeq 100 {\rm GeV}^2$
we obtain $R = 1.28$ as compared to $R = 1.47$ of ref.\ \cite{FFGS}.  Our study
of the off-diagonal effects of the input indicates that this is not the origin
of the difference.\footnote{We have an even larger difference, both numerically
and in the $x$ behaviour, with the results shown in Fig.~2 of 
ref.\ \cite{FFGS}.} \\
 
\noindent{\large \bf 6.~Conclusions} \\

We have studied the effect of using an off-diagonal gluon distribution 
$g (x, \: x^\prime, \: Q^2)$ to describe diffractive vector meson production.
As compared to the conventional diagonal gluon distribution we find
\be
\label{eq:b6}
R \; \equiv \; \frac{x^\prime g (x, \: x^\prime, \: Q^2)}{x g (x, Q^2)} \;
\simeq \; 1.1 - 1.4
\ee
depending on the kinematic variables, see Fig.~3.  In particular, $R$ increases
with $Q^2$ and with $\delta x = x - x^\prime$.  The enhancement is important
phenomenologically since the cross section for diffractive vector meson 
production depends on the {\it square} of the gluon distribution.  For example
for $J/\psi$ photoproduction, where the effective $Q^2 = M_{J/\psi}^2 / 4
\simeq 2.5 {\rm GeV}^2$ \cite{RRML}, there is an enhancement by a factor 1.2 
using off-diagonal evolution from $Q_0^2 = 1.5 {\rm GeV}^2$ with either the GRV
\cite{GRV} or the MRS(R2) \cite{MRS} gluon as input.  This is mainly a normalization effect and 
does not change the dependence of the cross section on the $\gamma p$
centre-of-mass energy $W$ (or on $x$), since $R$ mainly depends on the ratio 
$\delta x / x$.

In the absence of data, in Section~5, we studied a range of possibilities for
the input distribution $g (x, x^\prime, Q_0^2)$.  We find that the choice of 
input has little effect on the value of $R$ at large $Q^2$, but that it is not
negligible for $Q^2 \sim 10 {\rm GeV}^2$.  We therefore await data for the
determination of the input distribution. From this point of view the measurement
of $J/\psi$ production in a wider kinematic range is particularly attractive. As
there appears to be little intrinsic charm in the proton, we may hope to use
these data at a wide range of $W$ to fix the gluon input over an appreciable range
of $x$. \\

\noindent{\large \bf Acknowledgements} \\

MGR thanks the Royal Society, INTAS (95-311) and the Russian Fund of Fundamental
Research (96 02 17994), for support.


\newpage

\begin{figure}[htb]
   \vspace*{-1cm}
    \centerline{
     \psfig{figure=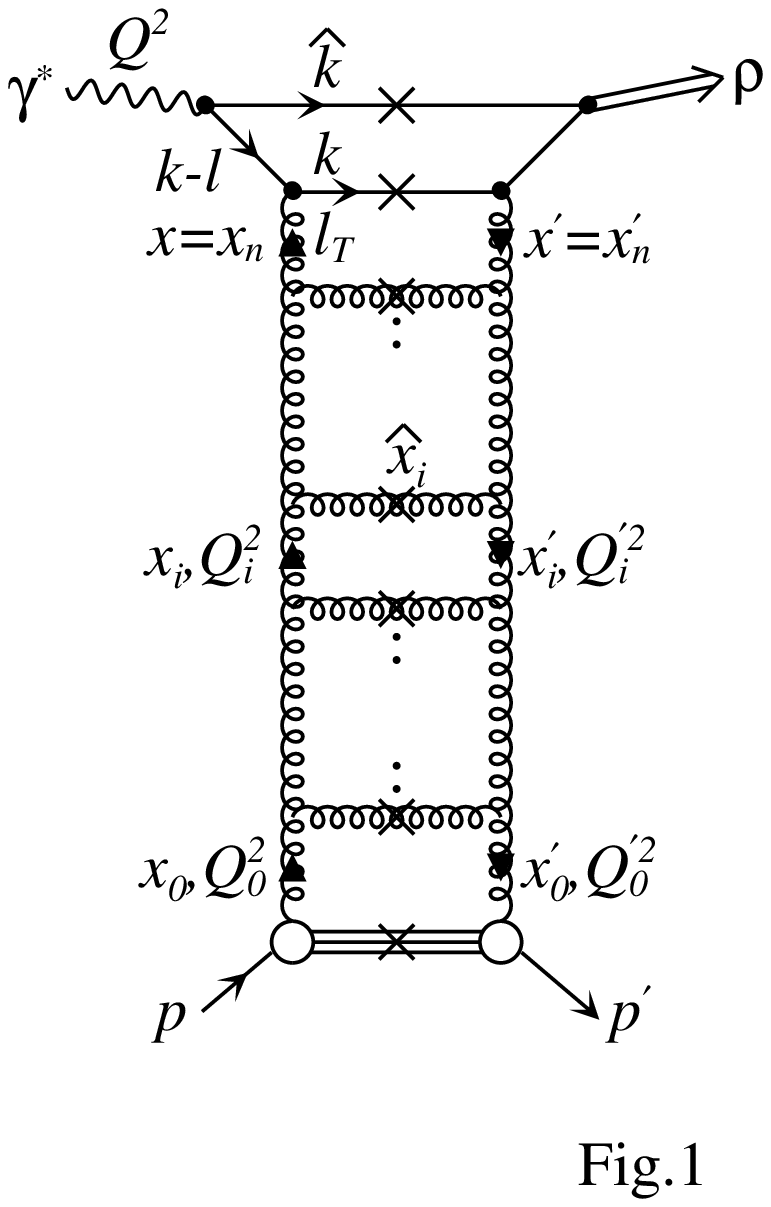,height=15cm,width=10cm}
               }
    \vspace*{-0.5cm}
     \caption{A diagrammatic representation of diffractive $\rho$ meson 
production, $\gamma^* p \rightarrow \rho p$, via a two-gluon exchange ladder. The 
crosses indicate that the particles are on-mass-shell in the calculation of the
imaginary part of the amplitude.
}
\end{figure}
\newpage

\begin{figure}[htb]
   \vspace*{-1cm}
    \centerline{
     \psfig{figure=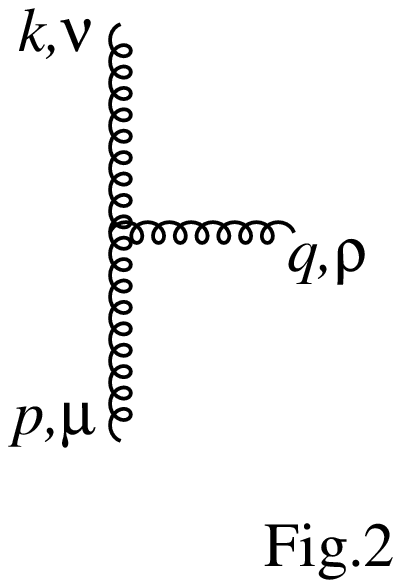,height=10cm,width=10cm}
               }
    \vspace*{-0.5cm}
     \caption{The variables of the three-gluon vertex used in the vertex function
of (\ref{eq:a24}).
}
\end{figure}
\newpage

\begin{figure}[htb]
   \vspace*{-1cm}
    \centerline{
     \psfig{figure=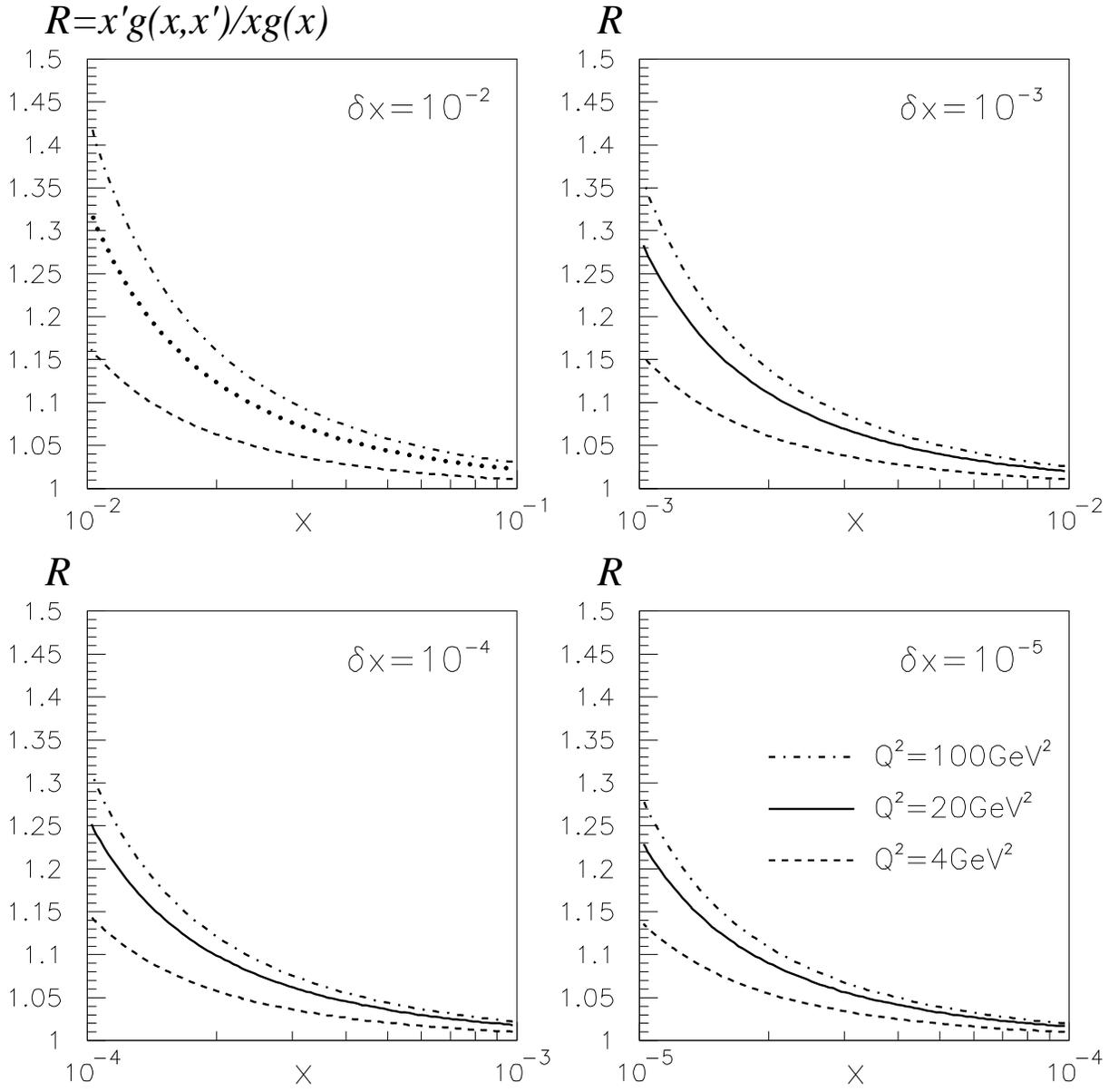,height=18cm,width=18cm}
               }
    \vspace*{-0.5cm}
     \caption{The ratio $R$ of the off-diagonal to diagonal gluon density as a 
function of $x$ for four values of $\delta x = x - x^\prime$ and three values of
$Q^2$, obtained from evolving from the GRV gluon \cite{GRV} with $R_0 = 1$ at
$Q_0^2 = 1.5 {\rm GeV}^2$.
}
\end{figure}
\newpage

\begin{figure}[htb]
   \vspace*{-1cm}
    \centerline{
     \psfig{figure=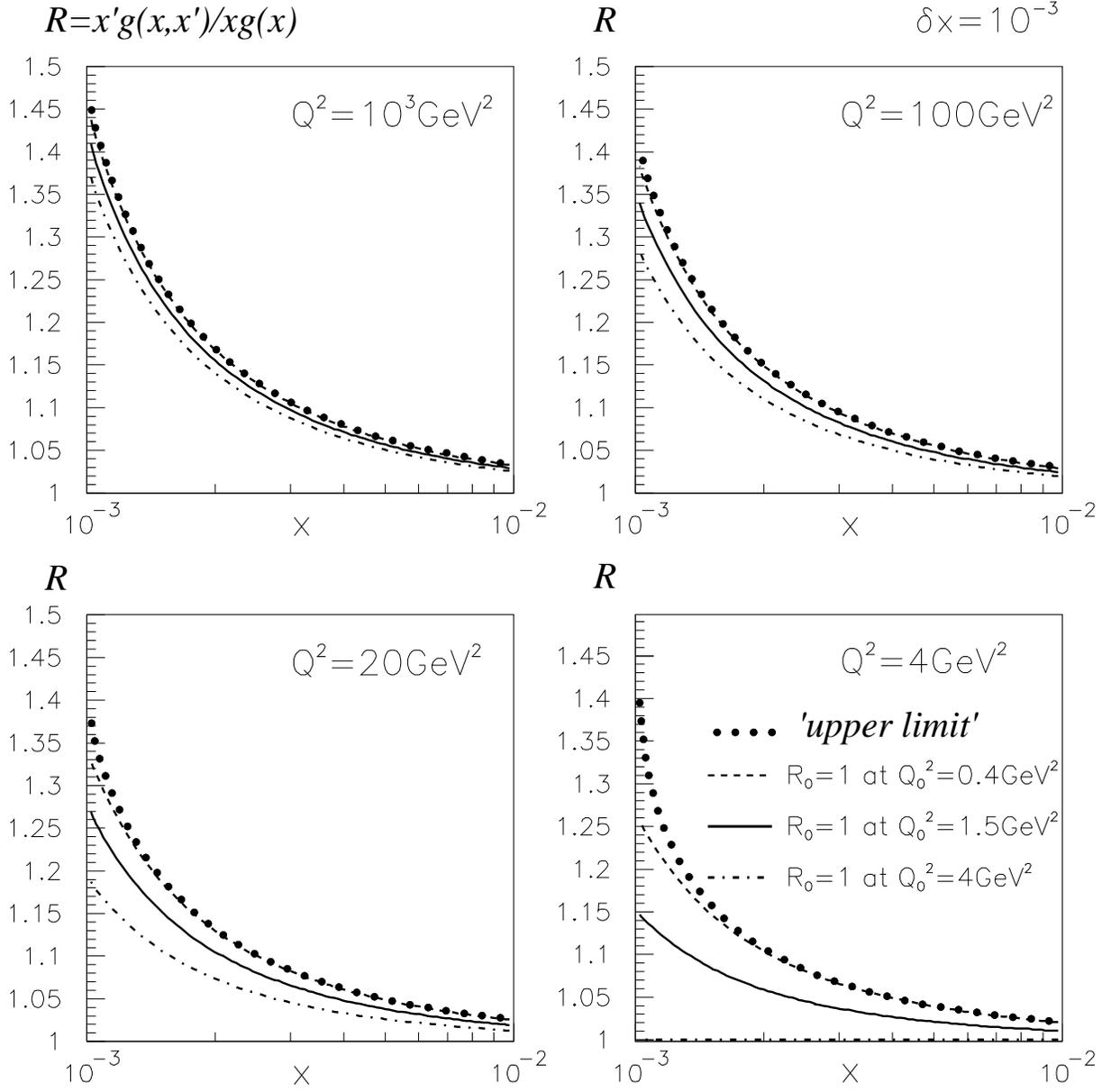,height=18cm,width=18cm}
               }
    \vspace*{-0.5cm}
     \caption{The dependence of the ratio of the off-diagonal to diagonal gluon
density on the input distribution. The dashed, continuous and dot-dashed curves
are obtained by evolving from $R_0 = 1$ at $Q_0^2 = 0.4, \: 1.5$ and $4 {\rm GeV}^2$, 
respectively, using the GRV gluon \cite{GRV}.  The upper limits (represented by 
dots) are obtained by evolving from $R_0$(max) of (\ref{eq:b3}) evaluated at 
$Q_0^2 = 1.5 {\rm GeV}^2$. The predictions corresponding to $Q_0^2 = 4 {\rm GeV}^2$
are shown only for comparison and should not be included in a realistic estimate
of the uncertainty in $R$.
}
\end{figure}
\newpage

\begin{figure}[htb]
   \vspace*{-1cm}
    \centerline{
     \psfig{figure=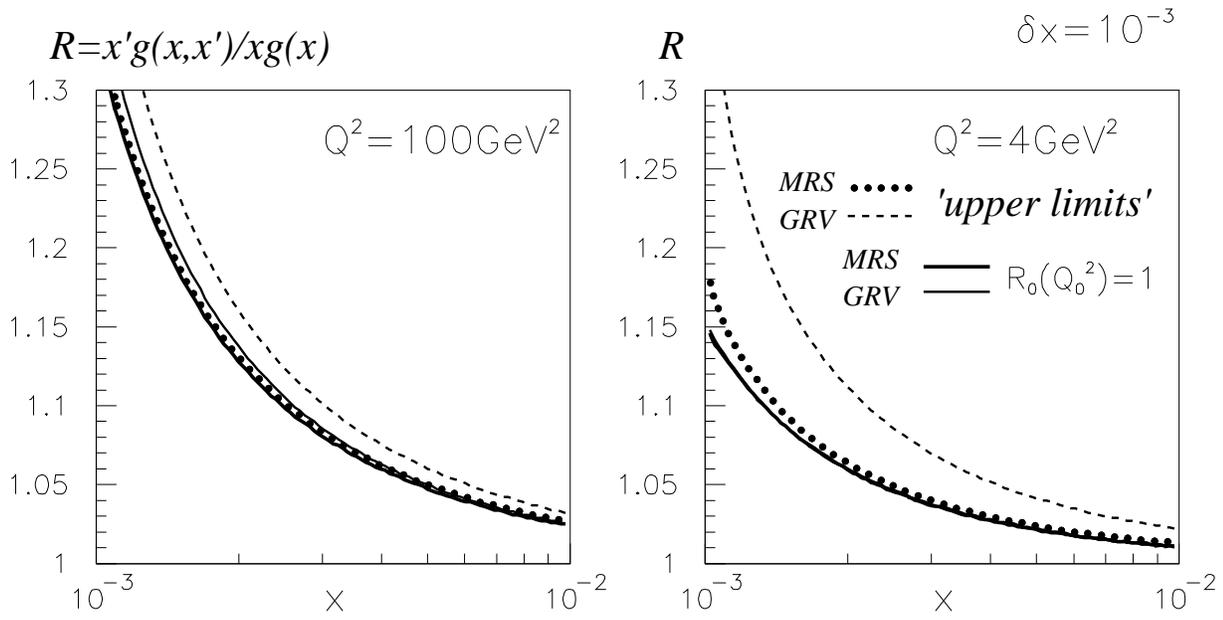,height=18cm,width=18cm}
               }
    \vspace*{-0.5cm}
     \caption{The predictions, together with the upper limits, based on evolution
from the MRS(R2) gluon \cite{MRS} compared with those obtained by evolving 
from the GRV gluon \cite{GRV}.  In each case we choose 
$Q_0^2 = 1.5 {\rm GeV}^2$.
}
\end{figure}

\end{document}